\documentclass[twocolumn,aps,amssymb]{revtex4}
\usepackage{graphicx}

\begin{document}
\title{
Terahertz Conductivity at the Verwey Transition in Magnetite }
\author{A. Pimenov}
\author{S. Tachos}
\author{T. Rudolf}
\author{A. Loidl}
\affiliation{Experimentalphysik V, EKM, Universit\"{a}t
Augsburg,
86135 Augsburg, Germany} %
\author{D. Schrupp}
\author{M. Sing}
\author{R. Claessen}
\affiliation{Experimentalphysik II, Universit\"{a}t
Augsburg, 86135 Augsburg, Germany}
\author{V. A. M. Brabers}
\affiliation{Department of Physics, Eindhoven University of
Technology, NL-5600, MB Eindhoven, The Netherlands}

\date{\today}

\begin{abstract}
The complex conductivity at the (Verwey) metal-insulator
transition in Fe$_3$O$_4$ has been investigated at THz and
infrared frequencies. In the insulating state, both the dynamic
conductivity and the dielectric constant reveal a power-law
frequency dependence, the characteristic feature of hopping
conduction of localized charge carriers. The hopping process is
limited to low frequencies only, and a cutoff frequency $\nu_1
\simeq 8\,$meV must be introduced for a self-consistent
description. On heating through the Verwey transition the
low-frequency dielectric constant abruptly decreases and becomes
negative. Together with the conductivity spectra this indicates a
formation of a narrow Drude-peak with a characteristic scattering
rate of about 5\,meV containing only a small fraction of the
available charge carriers. The spectra can be explained assuming
the transformation of the spectral weight from the hopping process
to the free-carrier conductivity. These results support an
interpretation of Verwey transition in magnetite as an
insulator-semiconductor transition with structure-induced changes
in activation energy.
\end{abstract}


\maketitle \section{Introduction}

Magnetite (Fe$_3$O$_4$) is probably the oldest known magnetic
material, which can be found in natural form. This material is of
considerable importance because of various applications in
magnetic recording and for high-frequency electronic devices.
Physical properties of magnetite have attracted much attention
after the discovery of the first-order metal-insulator transition
at $T_{\rm V}\sim 120\,$K. On cooling through the transition
temperature the dc-conductivity drops by two orders of magnitude.
A realistic model to explain the mechanism of this transition has
been suggested by Verwey \cite{verwey}, assuming a charge
order-disorder transition with alternating valence
(Fe$^{2+}/$\,Fe$^{3+}$) of the octahedrally-coordinated iron ions.
The metal-to-insulator transition in magnetite has been termed
Verwey transition since then. In spite of a large number of
experimental and theoretical efforts, the mechanism governing the
conduction and magnetic properties in magnetite is still under
debate \cite{walz,garsia}. According to the results of recent
x-ray resonant scattering \cite{garsiax}, even the charge-ordering
below $T_{\rm V}$ has been questioned and the concept of an
itinerant magnet was considered instead. Also the importance of
the orbital degrees of freedom have recently been highlighted
\cite{leonov}.

Magnetite crystallizes in a cubic high-temperature structure and
exhibits a monoclinic distortion below the Verwey transition. In
spite of enormous progress in resolving the low-temperature
monoclinic structure of magnetite \cite{iizumi,zuo,wright}, full
details could not be completely resolved so far. The monoclinic
phase is insulating and is characterized by a gap of the order of
$~100\,$meV in the excitation spectrum, which is seen in
photoemission data \cite{chainani,park,claessen}, in
thermoelectric properties \cite{knipers}, and by optical
spectroscopy \cite{tokura,gasp}. The charge transport in magnetite
is usually explained within a polaronic picture
\cite{tokura,gasp,leo,ihle}. However, there is no general
agreement about the energy of the polaronic absorption in the
conductivity spectra \cite{tokura,gasp,leo,ihle}. Recent
photoemission results \cite{claessen}, obtained in the same
magnetite crystals as in the present work, have been
self-consistently explained using a small-polaron model. A similar
concept has been applied recently to explain the dynamic
conductivity at the metal-insulator transition in quasi-one-
dimensional $\beta$-Na$_{0.33}$V$_2$O$_5$ \cite{marel}.

At low-temperatures the conduction in magnetite takes place via
hopping between localized states, which agrees well with both
activated dc-resistivity \cite{dc,matsui,kuipers} and with an
observed characteristic power-law $\sigma \propto \nu^s$
\cite{akishige} of the ac-conductivity. This universal
\cite{jonsher} power-law  has been observed in various materials
with internal disorder and is a clear fingerprint of hopping of
charge carriers \cite{long,elliott,dyre}. The situation is getting
more complicated in the metallic state. In spite of two
orders-of-magnitude increase in resistance above $T_{\rm V}$, the
absolute value of the dc-conductivity ($\sigma_{dc}(T \gtrsim
T_{\rm V}) \sim 50\,\Omega^{-1}$cm$^{-1}$) is still much smaller
than the Ioffe-Regel \cite{ioffe} minimum metallic conductivity
$\sigma_{min} \sim e^2/3\hbar a \sim 3000\,\Omega^{-1}$cm$^{-1}$,
even including the Mott correction \cite{mott1} in case of
disorder $\sigma_{min} \sim 0.03e^2/\hbar a \sim
300\,\Omega^{-1}$cm$^{-1}$ (here $a \sim 3\,$\AA \ is the
interatomic distance). The high-temperature conductivity of
magnetite is thus typical for a semiconductor with a high mobility
of thermally-excited charge carriers rather than for a metal.
Reviews of the recent and past developments in magnetite can be
found in \cite{walz,garsia,honig,tsuda,imada}.

In this paper we investigate Terahertz and infrared conductivity
of magnetite on both sides of the metal-to-insulator transition.
The unexpected result of these experiments is the abrupt change of
the dielectric constant at the transition temperature yielding
even negative values above $T_{\rm V}$. Together with the
conductivity spectra these results indicate the formation of a
narrow band of quasi-free carriers which contains only a small
part of the total spectral weight.

\section{Experimental details}

Synthetic single crystals of magnetite were prepared from
$\alpha-$Fe$_2$O$_3$ (spec. pure) using a floating-zone method
with radiation heating \cite{brabers}. Two single crystals were
used for the measurements, which revealed slightly different
transition temperatures, $T_{\rm V} \simeq 123\pm 1\,$K and
$116\pm 1$\,K, respectively. The transition temperature in
magnetite can be taken as a criterium for impurity concentration
and oxygen stoichiometry \cite{walz,impur} which, therefore,
differ significantly for the two crystals investigated. Both
samples revealed qualitatively similar electrodynamic properties
revealing only differences in absolute values. Therefore, in the
following mainly the results on a sample with $T_{\rm V} \simeq
123\,$K will be shown. Plane-parallel samples of different
thickness between 0.5\,mm and 0.05\,mm have been prepared from the
original boules. To ensure mechanical stability, the thinnest
samples were glued onto a MgO substrate of $\simeq 0.5\,$mm
thickness. The availability of samples with different thicknesses
is essential for transmittance experiments due to the strongly
temperature-dependent absorption in magnetite.

The dynamic experiments for frequencies $4$\,cm$^{-1}<\nu
<40$\,cm$^{-1}$ were carried out in a Mach-Zehnder interferometer
arrangement \cite{volkov} which allows both, the measurements of
the transmittance and the phase shift of a plane-parallel sample.
The complex transmission coefficient has been analyzed using the
Fresnel optical formulas for transmittance $T=|t|^2$
 of a plane-parallel sample \cite{born,heavens}:
\begin{eqnarray}
\label{eqtran}
\lefteqn{\hspace{2cm} t=\sqrt{T}e^{i \phi_T}=\frac{(1-r^{2})t_{1}}{1-r^{2}t_{1}^{2}} \ , }\\
& & \text{where\ }r=\frac{\sqrt{\varepsilon }-1}{\sqrt{\varepsilon
}+1}\ \text{ and } t_{1}=\exp (-2\pi i\sqrt{\varepsilon
}\,d/\lambda) \ \text{.}\nonumber
\end{eqnarray}
Here $r$ is the reflection amplitude of a thick sample, $t_1$ is
the ``pure''  transmission amplitude, $\varepsilon$ is the complex
dielectric permittivity
 of the sample, $d$ is the sample
thickness, and $\lambda$ is the radiation wavelength. It has been
assumed that the magnetic permeability $\mu=1$ in the frequency
range of our experiments. The transmittance for a two-layer system
can be obtained in a similar way  \cite{born,heavens}. Using Eq.\,
(\ref{eqtran}) the absolute values of the complex conductivity
$\sigma ^{*}=\sigma _{1}+i\sigma _{2}$ and dielectric constant
$\varepsilon^*=\varepsilon_1+i\varepsilon_2=\sigma^*/i\varepsilon_0\omega$
can be determined directly from the measured spectra. Here
$\varepsilon_0$ and $\omega=2\pi\nu$ are the permittivity of
vacuum and the angular frequency, respectively. In general, the
phase shift $\phi_T$ in the transmittance experiment contains a
possible ambiguity of $2\pi\cdot n$, where $n$ is an integer. To
exclude this source of errors, the reflectance of thick sample has
been measured in addition to transmittance experiments. A similar
Mach-Zehnder arrangement allowed to obtain the amplitude and the
phase shift of the reflectance at the Verwey transition. In total,
a set of four measured quantities was available to calculate the
real and imaginary part of the dynamic conductivity in the THz
frequency range.

In the infrared and visible frequency ranges 30\,cm$^{-1}<\nu
<21000$\,cm$^{-1}$ the complex conductivity has been obtained via
the Kramers-Kronig analysis of the reflectivity of the thick
sample. The reflectivity experiment were performed using Bruker
IFS-113v and IFS 66v/S Fourier-transform spectrometers. Different
sources, beam splitters and optical windows allowed to cover the
complete frequency range. In addition, the reflectance for the
frequency range $4<\nu <40$\,cm$^{-1}$ has been calculated using
the complex conductivity data, obtained by the transmittance
technique, described above. The combination of the results from
two experimental techniques \cite{mgb2} substantially expands the
low-frequency limit of the available spectrum and the quality of
the subsequent Kramers-Kronig transformation.

\section{Results and Discussion}

\begin{figure}[]
\centering
\includegraphics[width=8cm,clip]{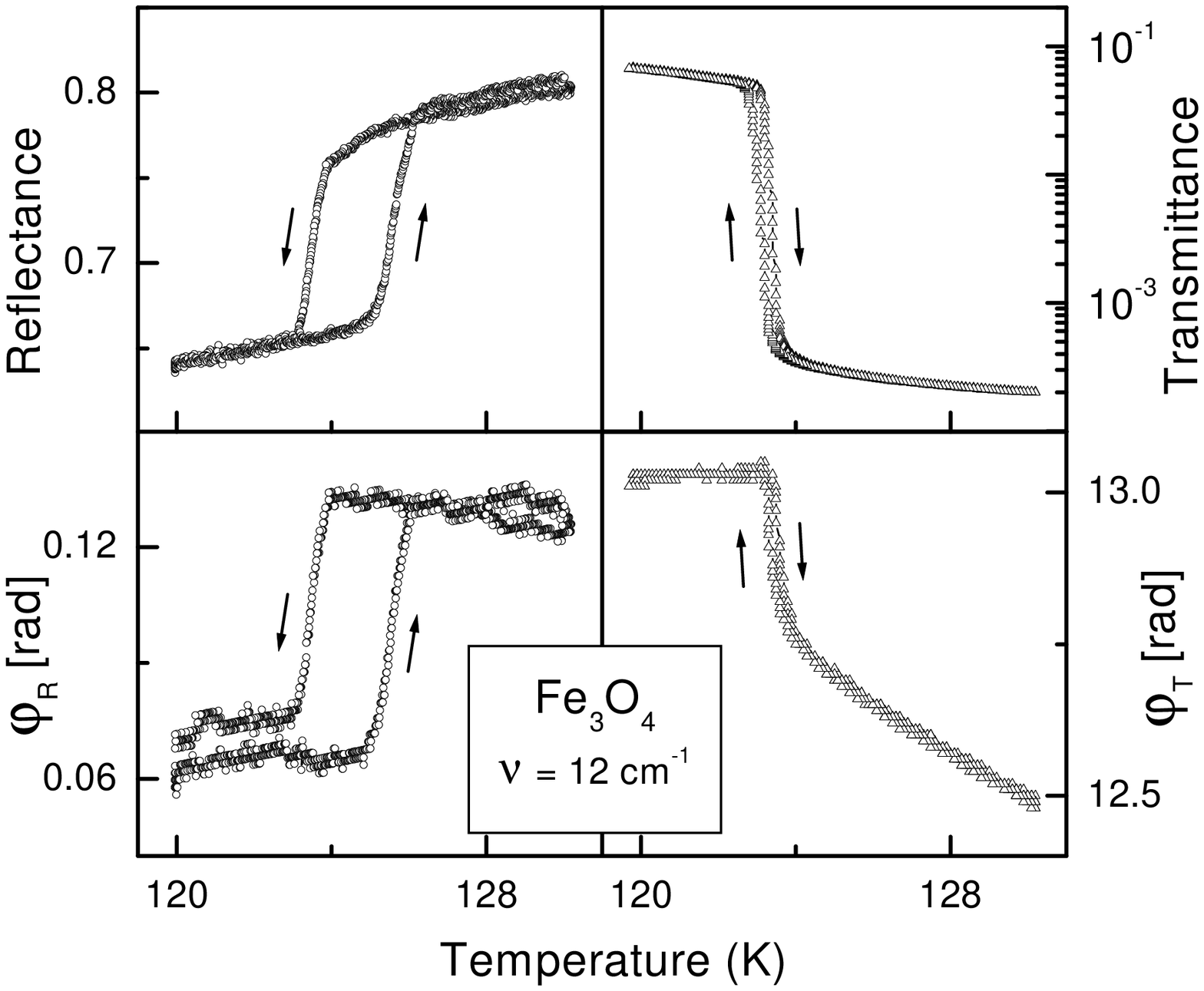}
%\vspace{0.2cm}
\caption{Left panels: temperature-dependent reflectance (top left)
and reflectance phase-shift (bottom left) of magnetite at
$\nu=12\,$cm$^{-1}$ in the vicinity of the Verwey transition.
Right panels: transmittance (top) and phase shift (bottom) of a
0.05\,mm thick plane-parallel sample glued on the 0.5\,mm thick
MgO substrate. The four measured quantities represent two
data-sets to determine $\sigma_1$ and
$\varepsilon_1=-\sigma_2/\varepsilon_0 \omega$. } \label{frefl}
\end{figure}

\begin{figure}[]
\centering
\includegraphics[width=8cm,clip]{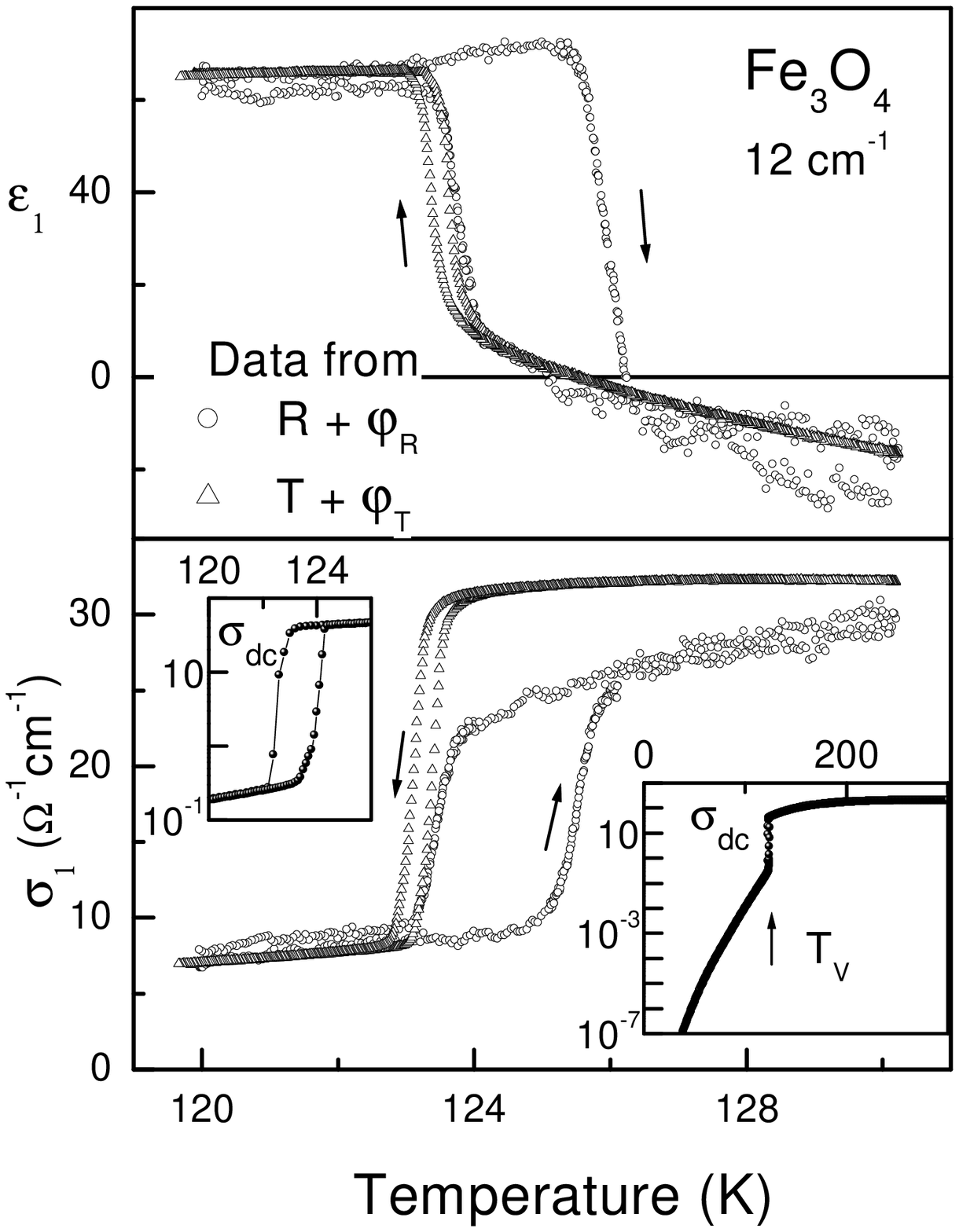}
%\vspace{0.2cm}
\caption{Temperature dependence of the dielectric constant
($\varepsilon_1$, upper panel) and conductivity ($\sigma_1$, lower
panel) in magnetite at $\nu = 12\,$cm$^{-1}$. The data have been
calculated from the complex reflectance (circles) of a thick
sample and from the complex transmittance of a 0.05\,mm thin plate
(triangles). Right inset - dc-conductivity of a thick sample in
full temperature range, left inset - temperature region close to
the Verwey transition.} \label{fsigt}
\end{figure}

Figure \ref{frefl} shows the complex reflectance and transmittance
of magnetite in the vicinity of Verwey transition. All four
quantities reveal strong changes at the transition temperature.
The most dramatic changes are observed in the transmittance, that
decreases by three orders of magnitude. Both transmittance and
reflectance show a hysteretic behavior, characteristic for a
first-order phase transition. However, the experimentally observed
hysteresis extends over different temperature regimes: the
hysteresis spans $\Delta T \simeq 2\,$K in the reflectance
experiments, and is significantly less than 0.5\,K in
transmittance experiments. We suggest that this distinct
difference probably results from the sample geometries, namely
utilizing a thick sample ($\simeq 5\,$mm) in the reflectance and
thin sample ($\simeq 0.05\,$mm) in the transmittance experiments.
Except for the hysteresis region, all measurements revealed
results coinciding within the experimental accuracy
(Fig.\,\ref{fsigt}). The effective over-determination of the
physical quantities, $\varepsilon_1$ and $\sigma_1$, in these
experiments was necessary because of the highly unusual
temperature and frequency dependence of the electrodynamic
properties of magnetite (as discussed below).

Figure \ref{fsigt} shows the temperature dependence of the
conductivity ($\sigma_1$) and the dielectric constant
($\varepsilon_1$) of magnetite at $\nu=12\,$cm$^{-1}$. The data
have been obtained either from the complex transmittance
($t=\sqrt{T}\exp(i\phi_T )$, triangles) or from the complex
reflectance ($r=\sqrt{R}\exp(i\phi_R )$, circles). The
conductivity reveals a sharp increase by a factor of three at the
transition temperature. However, this value is substantially lower
than the increase by two orders of magnitude, observed in the
dc-conductivity (insets of Fig.\,\ref{fsigt}). This is a direct
consequence of the power-law frequency dependence of the
ac-conductivity yielding orders-of-magnitude increase of the
conductivity with increasing frequency.

The behavior of the dielectric constant at the Verwey transition
is highly unusual. Both transmittance and reflectance experiments
lead to a strong \emph{decrease} of the dielectric constant at the
metal-to-insulator transition, becoming even negative above
$T\rm{_V}$. In the classical systems revealing a
metal-to-insulator transition (e.g. doped silicon) it has been
observed that the dielectric constant diverges approaching the
transition temperature from the insulating side \cite{mott1,hess},
which has been termed \emph{dielectric catastrophe}. In agreement
with this picture and using the same technique as presented here,
a strong increase of the dielectric constant at the
metal-insulator transition has been observed in Sr-doped LaMnO$_3$
\cite{lsmo} and recently in a magnetic field-induced transition in
Pr(Ca:Sr)MnO$_3$ \cite{pcsmo}. We believe that the first-order
character of the phase transition in magnetite interrupts the
increase of the dielectric constant, observed at $T<T_{\rm V}$ ($d
\varepsilon /d T \simeq 1\,$K$^{-1}$ at low frequencies). Instead,
the dielectric constant of magnetite abruptly jumps to negative
values at the "metallic" side of the transition.

\begin{figure}[]
\centering
\includegraphics[width=8cm,clip]{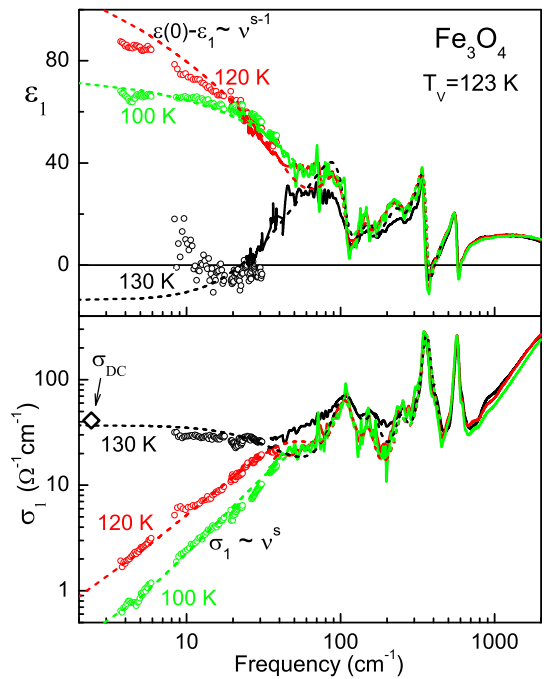}
%\vspace{0.2cm}
\caption{(color) Frequency dependence of the conductivity
($\sigma_1$, lower panel) and the dielectric constant
($\varepsilon_1$, upper panel) of magnetite above and below
$T_{\rm V} \simeq 123\,$K. Symbols below 40\,cm$^{-1}$ and solid
lines above 30\,cm$^{-1}$ represent the experimental data, dashed
lines - model calculations. Experimental data have been obtained
from transmittance and phase-shift of a thin Fe$_3$O$_4$ plate
below 40\,cm$^{-1}$, and via the Kramers-Kronig analysis of the
reflectance of the thick sample above 30\,cm$^{-1}$. The main
contribution to the model calculations is given by the hopping
term ($\sigma^* \propto \nu^s$, Eq.\,\ref{equn}) below $T_{\rm V}$
and a Drude-term (Eq.\,\ref{drude}) above $T_{\rm V}$.}
\label{fsigfr}
\end{figure}

Figure \ref{fsigfr} shows the frequency dependence of the
conductivity and dielectric constant in magnetite. Symbols at low
frequencies represent the results of the transmittance
experiments. Solid lines above 30\,cm$^{-1}$ have been obtained
via the Kramers-Kronig analysis of the reflectance. The lower
frame of Fig.\,\ref{fsigfr} shows the frequency dependence of
$\sigma_1$ above and below the Verwey transition. The
low-frequency conductivity in the insulating state is dominated by
a power-law in frequency ($\sigma_1 \propto \nu^s$ with $s\sim
1.3$) with a temperature-dependent amplitude and a weakly
temperature-dependent frequency exponent. As mentioned in the
Introduction, a power-law behavior of the conductivity is a
characteristic feature of hopping conduction between localized
states \cite{long,elliott,dyre}. Previously, the frequency
exponent $s\sim 0.7$ has been observed in magnetite at low
temperatures and at kHz-frequencies \cite{akishige}, a value
typical for the ac-conductivity in the audio-frequency range.
Substantially higher values of the exponent are expected at higher
frequencies: approaching the phonon-assisted-hopping regime,
$\sigma_1 \propto \nu^2$ has been predicted for low temperatures
\cite{mott}. The transition to higher frequency-exponents in the
microwave frequency range has been discussed recently for doped
semiconductors \cite{lee}. The power-law term is generally
referred to as \emph{universal dielectric response} \cite{jonsher}
and has been a recent matter of discussion concerning a possible
universal superlinear power law with $s \gtrsim 1$ at high
frequencies \cite{peter}.

The frequency dependence of the dielectric constant in magnetite
is shown in the upper panel of Fig.\,\ref{fsigfr}. The behavior of
the dielectric constant in the insulating state agrees
qualitatively with the power-law of $\sigma_1$, i.e
$\varepsilon_1$ increases with decreasing frequency. For $s>1$ the
low-frequency behavior can be approximated by
$\varepsilon_1(0)-\varepsilon_1(\nu) \propto \nu^{s-1}$.

Dashed lines in Fig.\,\ref{fsigfr} for the insulating state
($T<T_{\rm V}$) were obtained by fitting the experimental data to
the expression
\begin{equation}
\sigma_1(\nu)=\sigma_{dc}+ A \frac{(\nu/\nu_1)^s}{1+(\nu/\nu_1)^4}
\quad . \label{equn}
\end{equation}
Here $\sigma_{dc}$ is the dc-conductivity and $A$ is the amplitude
of the hopping process. Compared to the conventional form of this
response \cite{jonsher,long,elliott,dyre,peter}, we introduced an
additional high-frequency cutoff  $[1+(\nu/\nu_1)^4]$. The cutoff
frequency $\nu_1$ only weakly influences the conductivity in the
low-frequency range but prevents the divergence of the spectral
weight of $\sigma_1 \propto \nu ^s$ at high frequencies.

The analytical expression for the dielectric constant has been
obtained by applying the Kramers-Kronig transformation
\cite{wooten} to Eq.\,(\ref{equn}) which gives the following
analytical expression for the dielectric constant:
\begin{eqnarray}\label{eqeps}
\varepsilon_1(\nu)  = \frac{\sigma_1(\nu) - \sigma_{dc}}{ 2 \pi
\nu
\varepsilon_0} \Bigg\{ \tan(s\pi/2)- \qquad \nonumber\\
 - \frac{(\nu/\nu_1)^{1-s}}{\cos(s\pi/2)}
 \left[\sin[(s + 1)\frac{\pi}{4}]+\left(\frac{\nu}{\nu_1}\right)^2\cos[(s+1)\frac{\pi}{4}]\right]\Bigg\}
 \
\end{eqnarray}

This expression is valid both for $s<1$ and for $s>1$.

To take into account the phononic and high-frequency electronic
contributions, the sum of 6 Lorentzians has been added to
expressions Eq.\,(\ref{equn}) and Eq.\,(\ref{eqeps}). Five
Lorentzians at frequencies of 106, 152, 247, 361, and
574\,cm$^{-1}$ represent  phonon contributions. Additional
over-damped oscillator at 4270\,cm$^{-1}$ (0.53\,eV) approximates
the first excitation of the electronic origin. The given
characteristic frequencies agrees well with the published data
\cite{leo,tokura,gasp}. In order to simplify the analysis of the
THz conductivity and obtain the essential physics of the problem,
the contributions from the Lorentzians were fixed for all three
temperatures in Fig.\,\ref{fsigfr}. The parameters obtained from
simultaneous fitting of the conductivity and dielectric constant
are given in Table\,\ref{tab}.

\begin{table}
\caption{Parameters of the THz-frequency excitations in magnetite
close to the Verwey transition which correspond to
Eqs.\,(\ref{equn},\ref{eqeps},\ref{drude}). The changes in
conductivity can be well described assuming the transformation of
the hopping term Eq.\,(\ref{equn}) into the Drude-term
Eq.\,(\ref{drude}) with a high mobility of the quasi-free
carriers.} \label{tab} \vspace{0.5cm}
\begin{tabular}{cccc}
  \hline \hline
 T (K) & $\nu _{1}($cm$^{-1})$ & s &  $A(\Omega ^{-1}$cm$^{-1})$ \\
  \hline
  100 & 70 & 1.5 &  39 \\
  120 & 60 & 1.2 &  44 \\ \hline \hline
  T (K) & $1/2\pi \tau ($cm$^{-1})$ &  & $\sigma _{dc}(\Omega^{-1}$cm$^{-1})$
  \\ \hline
  130 & 40 &  & 37   \\
  \hline
\end{tabular}
\vspace{0.5cm}
\end{table}

A remarkable feature of the parameters for the low-temperature
insulating phase in Table\,\ref{tab} is the low value of the
cutoff frequency $\nu_1$. Indeed, a qualitative examination of the
conductivity frame of Fig.\,\ref{fsigfr} reveals that the hopping
process $\sigma_1 \propto \nu^s$ and the low-frequency edge of the
electronic excitation ($\nu> 1000\,$cm$^{-1}$) are well separated
in energy. This indicates that the cutoff frequency of the hopping
must be in the range $\sim 100\,$cm$^{-1}$, in agreement with the
fits. This result is rather surprising and suggests that the
low-temperature conductivity is governed by phonons.

In contrast to the gradual variation of $\varepsilon_1(\omega,T)$
and $\sigma_1(\omega,T)$ at low temperatures, dramatic changes are
observed at the Verwey transition. At low frequencies the
conductivity increases by more than one order-of-magnitude. This
increase is substantially smaller than a change of two
orders-of-magnitude in dc-conductivity (see insets of
Fig.\,\ref{fsigt}). This difference is the result of a power-law
frequency dependence of the conductivity in the insulating state.
As a consequence, for frequencies close to 35\,cm$^{-1}$ we
observe almost \emph{no conductivity change} at $T_{\rm V}$. This
observation has been verified in a separate temperature-dependent
experiment at $\nu=35\,$cm$^{-1}$ (not shown).

As discussed above, the decrease of the dielectric constant at the
transition to the conducting state is not expected within the
scope of a conventional metal-insulator transition scenario. In
order to find an explanation to this experimentally-observed
behavior, we mention that a negative dielectric constant naturally
follows by assuming a Drude-response of quasi-free carriers,
\begin{equation}
\sigma_{\rm DR}^* = \sigma_{dc}/(1-i 2 \pi \nu \tau) \ ,
\label{drude}
\end{equation}
where $\sigma_{dc}$ and $1/2\pi\tau$ are dc-conductivity and
scattering rate, respectively. In addition to the negative changes
in $\varepsilon_1$, the conductivity in the metallic state shows a
slight downward curvature in the frequency dependence (lower panel
of Fig.\,\ref{fsigfr}). This provides a further argument to
include a Drude-term to the model. Indeed, the simplest way to
reproduce the experimental changes between 120 and 130 K is just
to \emph{replace} the hopping term (Eq.\,\ref{equn},\ref{eqeps})
by the Drude term. For the second sample of magnetite with $T_{\rm
V} \simeq 116\,$K a qualitatively similar behavior has been
observed and the absolute values of the parameters agreed within
$\sim 20\%$. We believe that this variation partly corresponds to
the difference in  the conductivities of the samples.

The low value of the scattering rate, as observed in the present
experiments, corresponds to a high carriers mobility of
$\mu=e\tau/m\simeq10^3\,$cm$^2/Vs$ (assuming $m$ being the free
electron mass) and correlates well with the high quality of our
magnetite single crystals. Due to the low characteristic frequency
of the coherent (itinerant) process, its spectral weight
\begin{equation}
S=\frac{2}{\pi}\int\limits_{0}^{\infty}
\sigma_{1,DR}(\omega)d\omega = \sigma_{dc}/\tau
\end{equation}
is small compared to the full number of carriers. Taking as an
estimate of the total spectral weight the concentration of
Fe$^{2+}$ ions \cite{tsuda}, only about $2\cdot 10^{-4}$ of the
theoretically-available carriers participate in the coherent
process. We recall further, that within the rough estimate the
spectral weight of thermally-activated quasi-free carriers in a
semiconductor can be written as
\begin{equation}
% S=n\exp(-\frac{\Delta_m}{k_{\rm B}T})e^2/m_{eff} \ ,
 S=\frac{ne^2}{m_{eff}}\exp(-\frac{\Delta_m}{k_{\rm B}T}) \ ,
\end{equation}
where $n$, $e$ and $m_{eff}$ are concentration, charge and
effective mass of the charge carriers, respectively. In this
expression the effective concentration of charge carriers is
temperature-activated and is governed by an energy gap $\Delta_m$.
The estimate of the effective mass $m_{eff} \simeq 10^2 m$ has
been obtained from the infrared conductivity \cite{gasp} and from
the small-polaron analysis of photoemission of the same sample
\cite{claessen}. With this estimate for $m_{eff}$ the observed
small spectral weight of the Drude-process corresponds to
$\Delta_m \simeq 40\,$meV, which correlates with a
weakly-activated dc-conductivity just above $T_V$ (right inset in
Fig.\,\ref{fsigt}). Within the same picture, the Verwey transition
in magnetite corresponds just to an increase of the characteristic
activation energy at the insulating side of the transition. The
drop of dc-conductivity at the transition can be ascribed to a gap
change from $\Delta_m \simeq 40\,$meV (conducting state) to
$\Delta_i \simeq 90\,$meV (insulating state). This new value of
the energy gap agrees with the activation energy of
dc-conductivity \cite{dc,matsui,kuipers}, thermoelectric power
\cite{knipers}, and with the gap values obtained in photoemission
experiments \cite{chainani,park,claessen}. The presented
description implies the interpretation of Verwey transition as an
insulator-semiconductor transition with a change in activation
energy induced by crystal structure or by electronic
configuration.

\section{Conclusions}

Terahertz and infrared conductivity in magnetite has been obtained
on both sides of Verwey metal-to-insulator transition. Two
orders-of-magnitude change in dc-conductivity halves at millimeter
frequencies and even disappears for $\nu\gtrsim 30\,$cm$^{-1}$. In
the far-infrared range the effect of the transition remains solely
in changing the dielectric constant ($\varepsilon_1$) of
magnetite. The change of $\varepsilon_1$ during the transition
into the conducting state is negative, which is in surprising
contrast to an expected (positive) divergence of the dielectric
constant at a metal-insulator transition. Within the simple
model-analysis the conductivity mechanism switches  between
hopping of localized carries  below $T_{\rm V}$ and itinerant
motion above $T_{\rm V}$. These results evidence the formation of
the coherent state on the metallic side of the Verwey transition
with high mobility of the charge carriers. Together with the
recent photoemission data on the same samples the interpretation
of the Verwey transition as an insulator-semiconductor transition
is suggested.

\section{Acknowledgements}

The stimulating discussion with P. Lunkenheimer, L. V. Gasparov
and I. Leonov is gratefully acknowledged. We thank A. A. Volkov,
P. Pfalzer and A. Pimenova for help in sample preparation and S.
Klimm for the measurement of the dc-conductivity. This work was
supported by BMBF (13N6917/0 - EKM) and by DFG (SFB484 -
Augsburg).

\end{document}